\documentstyle[prl,aps,twocolumn]{revtex}
\begin{document}
\draft
\title{Josephson Current through Charge Density Waves}
\author{M. I. Visscher and B. Rejaei}
\address{Theoretical Physics Group, Department of Applied Physics/DIMES \\
Delft University of Technology, Lorentzweg 1, 2628 CJ Delft, The Netherlands}
\date{\today}
\maketitle

\begin{abstract}
The effect of the collective charge density wave (CDW) motion on the Josephson 
current in a superconductor/charge density wave/superconductor 
junction is studied theoretically. By deriving the kinetic equations for the coupled 
superconductor-CDW system, it is shown that below the critical current the CDW does not move. 
Biased above this value, the Josephson current oscillates as a function of the velocity of the 
sliding CDW and the collective mode acts as a non-linear shunting resistor parallel 
to the Josephson channel. The effects of impurity pinning are accounted for at a phenomenological level
\end{abstract}
\pacs{PACS numbers: 74.50, 72.15.N}

The Josephson effect\cite{josephson} is known to exist in superconductor
hybrid structures, where two superconductors are separated by insulating
barriers or normal metals \cite{kulik}. Recently, the theory of the
Josephson effect through systems which support a non-Fermi
(Tomonaga-Luttinger) liquid ground state has received much attention due to
rapid developments in fabrication technology \cite{odintsov}. In this Letter
we investigate the Josephson current through a different yet related system,
namely a strongly anisotropic metal with a charge density wave (CDW)
instability. Throughout this paper our attention will be restricted to incommensurate
CDW's. Despite the insulating quasi-particle spectrum, incommensurate 
CDW's allow for a unique collective mode of transport, which distinguishes them from
ordinary insulators \cite{gorkov}. The sliding motion of the CDW carries
electrical current and is dissipationless in the absence of damping sources.
It is therefore interesting to investigate whether the Josephson current in
a superconductor/charge density wave/superconductor (S/C/S) junction will be
affected by the sliding motion. Recent progress in controlled deposition of
thin films of CDW's may lead to the fabrication of such mesoscopic-scale
heterostructures in the near future \cite{zant}.

Our model is a current biased S/C/S junction, which consists of parallel
one-dimensional CDW chains of length $L$, sandwiched between two large
superconductors with phases $\varphi _R$ and $\varphi _L$ (Fig. 1). The
CDW is characterized by a complex order parameter $|\Delta
_c|e^{i\chi }$, where $|\Delta _c|$ is proportional to the amplitude of the
density modulation and the phase $\chi $ determines its position. The CDW
can slide along the chains with velocity proportional to $\dot \chi \equiv
\partial _t\chi $.

Using the Keldysh formalism for superconductors\cite{larkin} and CDW's 
\cite{artemenko} we have formulated a consistent framework for the
quasi-classical dynamics of the coupled superconductor-CDW system. It is
shown that the CDW is immobile under a certain critical current. Biased
above this value, the CDW starts to slide and the collective transport mode
causes a shunting resistance parallel to the Josephson channel. We show
that, in the quasi-stationary approximation, the Josephson current has an
oscillatory behavior as a function of the phase $\theta =\varphi _R-\varphi
_L+\dot \chi L/v_F,$ where $v_F$ is the Fermi velocity. Apparantly CDW motion
induces a dynamical phase which is added to the conventional superconductor
phase difference. Finally, we discuss the effects of impurity pinning at a
phenomenological level.

The dynamics of superconductor and CDW systems can be described
simultaneously by the semiclassical Green functions $g_{\alpha \beta
}^i(x;t,t^{\prime })$ where $i=\left\{ R,A,K\right\} $ and $\alpha ,\beta
=\left\{ 1,2,3,4\right\} .$ The retarded ${\bf g}^R$ and advanced functions ${\bf g}^A$ 
determine the excitation spectrum and the Keldysh function ${\bf g}^K
$ describes the kinetics of the system. The subscripts $1,3$ refer to right
and left moving electrons with spin up and $2,4$ to left and right moving
holes with spin down. Throughout this paper ``hat" denotes (2x2) matrices
and bold-face (4x4) matrices. The Green functions satisfy the equation of
motion 
\begin{equation}
\label{kinetic}i{\hbar }v_F\partial _x{\bf g}^i+\left[ {\bf H\stackrel{\circ 
}{,}g}^i\right] _{-}=0,
\end{equation}
where 
$$
{\bf H}=i{\hbar }\partial _t\bbox{\sigma}_3{\bf \Sigma }_3-\Phi {\bf \Sigma }%
_3+{\bf \Delta } 
$$
$$
\bbox{\sigma}_k=\left( 
\begin{array}{cc}
\sigma _k & 0 \\ 
0 & \sigma _k
\end{array}
\right) ,\quad {\bf \Delta }=\left( 
\begin{array}{cc}
\hat \Delta _s & -
\hat \Delta _c \\ \hat \Delta _c^{\dagger } & -\hat \Delta _s
\end{array}
\right) , 
$$
\begin{equation}
{\bf \Sigma }_1=\left( 
\begin{array}{cc}
0 & \hat 1 \\ \hat 1 & 0
\end{array}
\right) ,{\bf \Sigma }_2=i\left( 
\begin{array}{cc}
0 & -
\hat 1 \\ \hat 1 & 0
\end{array}
\right) ,{\bf \Sigma }_3=\left( 
\begin{array}{cc}
\hat 1 & 0 \\ 
0 & \hat 1
\end{array}
\right).
\end{equation}
Here $\Phi $ is the quasi-particle potential, $\sigma _k$ with 
$k=\left\{ 1,2,3\right\}$ are the three
Pauli matrices. The dot operation $\circ $
denotes internal time integrations as well as matrix multiplication and the
brackets $[\,]_{-}$ denote commutation. The self-energy term for impurity
scattering is neglected throughout this paper. 
The matrix $\hat \Delta _s$ is given by $\hat \Delta
_{s,11}=\hat \Delta _{s,22}=0,$ $\hat \Delta _{s,12}=-\hat \Delta
_{s,21}^{*}=|\Delta _s|\exp (i\varphi )$, and $\hat \Delta _c=\hat 1|\Delta
_c|\exp (i\chi )$. 

It is convenient to gauge away both phases $\varphi $ and
$\chi $ by applying the unitary transformation
\begin{equation}
\label{gauge}{\bf \tilde g}^i={\bf U}^{\dagger }(x,t)\circ {\bf g}^i\circ 
{\bf U}(x,t^{\prime }) 
\end{equation}
where ${\bf U}=\exp (\frac i2{\bf \Sigma }_3\chi +\frac i2\bbox{\sigma}%
_3\varphi ).$ Disregarding local variations of $\dot \chi $, we look
for a stationary state solution of the form ${\bf g(}x{\bf ,}t-t^{^{\prime }}%
{\bf ),}$ which can be treated by the Fourier transformation
\begin{equation}
\label{fourier}{\bf \tilde g}^i(x;t-t^{\prime })=\int \frac{d\epsilon }{2\pi 
}{\bf \tilde g}^i(x,\epsilon )e^{-i\epsilon (t-t^{\prime })/{\hbar }}.
\end{equation}
The stationary-state equation of motion for the Fourier transformed function is
\begin{equation}
\label{stationary}iv_F\partial _x{\bf \tilde g}^i+\left[ \epsilon 
\bbox{\sigma}_3{\bf \Sigma }_3-\tilde \Phi {\bf \Sigma }_3-ev_F\tilde A%
\bbox{\sigma}_3+i|{\bf \Delta }|,{\bf \tilde g}^i\right] _{-}=0.
\end{equation}
with $\tilde \Phi =\Phi +\frac 12{\hbar }v_F\partial _x\chi +\frac 12{\hbar }%
\dot \varphi $, $\tilde A={\hbar }\dot \chi /2ev_F+{\hbar }\partial
_x\varphi /2e$ and $|{\bf \Delta }|=|\Delta _s|\bbox{\sigma}_2{\bf \Sigma }%
_3-|\Delta _c|{\bf \Sigma }_2.$ From the structure of this equation the
well-known duality between the superconductor and CDW phases $v_F\partial
_x\chi \Leftrightarrow \dot \varphi $ and $\dot \chi \Leftrightarrow
v_F\partial _x\varphi $ is observed. The gradient of the superconductor
phase and the time derivative of the CDW phase correspond to an electrical
current, whereas the gradient of the CDW phase and the time derivative of
the superconductor phase correspond to an electronic potential.

For the inhomogeneous S/C/S system Eq. (\ref{stationary}) has to be
supplemented by boundary conditions, which adequately describe the two
superconductor interfaces. Here we restrict ourselves to the ideal case
where no defects or potential barriers are present at the interaces. In
order to derive the boundary conditions it will be convenient to decompose
the (4x4) Green functions into four (2x2) blocks as follows

\begin{equation}
{\bf g}=\left( 
\begin{array}{cc}
\hat g & \hat f \\ -{\hat{\bar f}} & -{\hat{\bar g}}
\end{array}
\right).
\end{equation}
It can be shown \cite{zaitzev} that the diagonal blocks are normalized as
usual 
\begin{equation}
\label{norm}\hat g\circ \hat g=\hat{\bar{g}}\circ \hat{\bar{g}}=\hat 1\delta
(t-t^{\prime }),
\end{equation}
and the non-diagonal blocks must satisfy the following relations in the
superconducting leads on the left (L) and right (R) 
\begin{mathletters}
\label{bc:all}
\begin{equation}
\label{bc:a}\hat g_L\circ \hat f_L=-\hat f_L,\quad \hat{\bar{g}}_L\circ 
\hat{\bar{f}}_L=\hat{\bar{f}}_L,
\end{equation}
\begin{equation}
\label{bc:b}\hat g_R\circ \hat f_R=\hat f_R,\quad \hat{\bar{g}}_R\circ 
\hat{\bar{f}}_R=-\hat{\bar{f}}_R.
\end{equation}
\end{mathletters}
The retarded and advanced Green functions in the superconductors are
determined from the stationary-state equation of motion Eq. (\ref{stationary}%
) with $\tilde \Phi =|\Delta _c|=0$. Assuming $\Delta _s$ to be constant in
the superconductors, it is convenient to apply the Bogoliubov transformation:
$$
\bbox{\cal{G}}=\bbox{\vartheta}^{-1}{\bf \tilde g}\bbox{\vartheta},
$$
\begin{equation}
\label{bogol}
{\bbox{\vartheta}}=\left( 
\begin{array}{cc}
\hat \vartheta _{+} & 0 \\ 
0 & \hat \vartheta _{-}
\end{array}
\right) ,\quad \hat \vartheta _{\pm }=\left( 
\begin{array}{cc}
u_{\pm } & -v_{\pm } \\ 
v_{\pm } & -u_{\pm }
\end{array}
\right) ,
\end{equation}
where $u_{\pm }$ and $v_{\pm }$ are the gauge transformed BCS coherence
factors given by
$$
u_{\pm }=\sqrt{\frac 12(1+\frac{\lambda _{\pm }}{\varepsilon _{\pm }})}\quad
v_{\pm }=-\sqrt{\frac 12(1-\frac{\lambda _{\pm }}{\varepsilon _{\pm }})},
$$
\begin{eqnarray}
\lambda _{\pm } &=& 
{\rm sign}( \varepsilon_{\pm} )\sqrt{\varepsilon_{\pm} ^2-|\Delta _s|^2} \: \Theta (|%
\varepsilon_{\pm} |-|\Delta _s|) \nonumber \\ & & + i\sqrt{|\Delta _s|^2-\varepsilon_{\pm} ^2}%
\: \Theta (|\Delta _s|-| \varepsilon_{\pm}|). 
\end{eqnarray}
Here $\varepsilon _{\pm }=\epsilon \mp {\hbar }\dot \chi /2,$ and $\Theta $ is
the Heavyside step-function. After substitution of Eq. (\ref{bogol}), it follows together with
conditions (\ref{norm}) that the
retarded Green functions $\hat{\cal{G}}$ and ${\hat{\bar{\cal{G}}}}$ at
the interfaces are
\begin{mathletters}
\label{bd:all}
\begin{equation}
\label{bd:a}\hat{\cal{G}}_L=\left( 
\begin{array}{cc}
1 & 0 \\ 
C_L & -1
\end{array}
\right) ,\quad \hat{\bar{\cal{G}}}_L=\left( 
\begin{array}{cc}
1 & \bar C_L \\ 0 & -1
\end{array}
\right) 
\end{equation}

\begin{equation}
\label{bd:b}\hat{\cal{G}}_R=\left( 
\begin{array}{cc}
1 & C_R \\ 
0 & -1
\end{array}
\right) ,\quad \hat{\bar{\cal{G}}}_R=\left( 
\begin{array}{cc}
1 & 0 \\ 
\bar C_R & -1
\end{array}
\right) 
\end{equation}
where the constants $C_{R,L}$ and $\bar C_{R,L}$ denote Andreev scattering
amplitudes for the scattering of an quasi-electron into a quasi-hole and
vice-versa. The $\hat{\cal{F}}$ and $\hat{\bar{\cal{F}}}$ blocks are given by

\begin{equation}
\label{bd:c}\hat{\cal{F}}_L=\left( 
\begin{array}{cc}
0 & 0 \\ 
D_{L1} & D_{L2}
\end{array}
\right) ,\quad \hat{\bar{\cal{F}}}_L=\left( 
\begin{array}{cc}
\bar D_{L1} & \bar D_{L2} \\ 0 & 0
\end{array}
\right) 
\end{equation}

\begin{equation}
\label{bd:d}\hat{\cal{F}}_R=\left( 
\begin{array}{cc}
D_{R1} & D_{R2} \\ 
0 & 0
\end{array}
\right) ,\quad \hat{\bar{\cal{F}}}_R=\left( 
\begin{array}{cc}
0 & 0 \\ 
\bar D_{R1} & \bar D_{R2}
\end{array}
\right),
\end{equation}
\end{mathletters}
where the diagonal $D$-constants are 'normal' backscattering amplitudes
without spin-flip ('normal' means non-Andreev), and the non-diagonal $D$%
-constants are scattering amplitudes with spin-flip, which will turn out to
be zero. The boundary conditions (\ref{bd:all}) state that both quasi-electrons and
quasi-holes which move away from the CDW region into the ideal
superconducting leads will never be reflected into quasi-particles moving in
the opposite direction \cite{rejaei}. The boundary conditions for the
advanced Green functions are obtained by the relation ${\bf g}^A=-\bbox{\sigma}%
_3({\bf g}^R)^{\dagger }\bbox{\sigma}_3$. 

In order to determine the yet 
unknown constants $C's$ \& $D's$, we have to solve Eq. (\ref{stationary}) in
the CDW region and match the boundary conditions. 
Neglecting the spatial variations of $\ |\Delta _c|$ near the contacts,
the solution to Eq. (\ref{stationary}) in the CDW region in terms of Bogoliubov
transformed functions is
$$
{\bbox{\cal{G}}}(x)={\bbox{\cal{M}}}^{-1}{\bbox{\cal{G}}}(0){\bbox{\cal{M}}},
$$
\begin{equation}
\label{solution}
\bbox{\cal{M}}=\bbox{\vartheta}^{-1}\exp(-\frac{i{\bf{H}}x}{\hbar v_{F}})\bbox{\vartheta},
\end{equation}
where
${\bf H}=\epsilon \bbox{\sigma}_3{\bf \Sigma }
_3-\frac{1}{2}\hbar {\dot{\chi}}\bbox{\sigma}_{3}-i|\Delta _c|{\bf \Sigma }_2$.
Evaluating $\bbox{\cal{G}}(x)$ at $x=L$ and inserting the boundary conditions 
(\ref{bd:a})-(\ref{bd:d}) we obtain the relation which connects the Green functions
at the left and right interface
\begin{equation}
\label{transcon}
\bbox{\cal{L}}\bbox{\cal{G}}_R{\bf =}\bbox{\cal{G}}_L
\bbox{\cal{L}},
\end{equation}
in terms of the transfer matrix $\bbox{\cal{L}}=\bbox{\cal{M}}(L)$:
$$
\bbox{\cal{L}}=\bbox{\vartheta}^{-1}\exp (\frac i2\theta 
\bbox{\sigma}_3)\left\{ \cos (\frac{\lambda _cL}{\hbar v_F}){\bf 1}-i\frac{%
{\bf \tilde H}}{\lambda _c}\sin (\frac{\lambda _cL}{\hbar v_F})\right\} {%
\bbox{\vartheta}},
$$
$$
\tilde{\bf H}=\epsilon \bbox{\sigma}_3{\bf \Sigma }
_3-i|\Delta _c|{\bf \Sigma }_2,
$$
\begin{eqnarray}
\label{transfer}
\lambda _c &=& \sqrt{\epsilon ^2-|\Delta _c|^2} \: \Theta (|\epsilon |-|\Delta _c|) \nonumber \\ 
& & +i\sqrt{|\Delta _c|^2-\epsilon ^2} \: \Theta (|\Delta _c|-|\epsilon |),
\end{eqnarray}
with $\theta =\varphi _R-\varphi _L+\dot \chi L/v_F.$ Evidently, the extra
phase factor arises from the line integral of the vector potential $\tilde A$
along the junction $\frac{2e}\hbar \int_0^Ldx\tilde A=\dot \chi L/v_F$. Since for $|\Delta _c|\neq 0,$
the solutions are not invariant under the gauge transformation, it is
expected from Eq. (\ref{transfer}) that in addition to $\varphi _R-\varphi _L
$ the dynamcial phase $\dot \chi L/v_F$ due to the motion of the CDW will
also enter the expression for the Josephson current. Inserting the boundary conditions, 
which describe the two
superconductors, we can solve for the scattering constants $C^{\prime }s$
and $D^{\prime }s$. Details will be given elsewhere \cite{visscher}. We next substitute the obtained
solutions into the expression for the total current $I$ through the S/C/S junction
\begin{equation}
\label{current}-I=\frac{ev_FN(0)}8\int d\epsilon {\rm Tr}\bbox{\sigma}_3{\bf 
\tilde g}^K+ev_FN(0){\hbar }\dot \chi .
\end{equation}
with $-e$ the electron charge and $N(0)=(\pi \hbar v_F)^{-1}$ the
density of states at the Fermi level for one spin direction. The result is
\begin{equation}
\label{result}I=-\frac e\pi \dot \chi +\frac{e}h{\rm Im}\int d\epsilon 
\frac{\sin \eta }{\cos \eta -\cos \zeta }h_0^{+}.
\end{equation}
In this expression $h_0^{+}=\tanh (\frac{\epsilon -\hbar \dot \chi /2}{2k_BT})+\tanh (%
\frac{\epsilon +\hbar \dot \chi /2}{2k_BT}),$ and we have defined the
complex angles 
\begin{equation}
\eta =\theta +\xi _{+}-\xi _{-},\quad \xi _{\pm }=\frac i2\ln \frac{%
\varepsilon _{\pm }-\lambda _{\pm }}{\varepsilon _{\pm }+\lambda _{\pm }},
\end{equation}
and 
\begin{eqnarray}
& &\cos \zeta = 1- \nonumber \\ 
& &2\left( \cos \frac{\lambda _cL}{\hbar v_F}\sin \frac {\xi_{+}+\xi_{-}}{2}-\frac 
\epsilon {\lambda _c}\sin \frac{\lambda _cL}{\hbar v_F}\cos \frac {\xi_{+} + 
\xi_{-}}{2}
\right) ^2.
\end{eqnarray}
In the static limit where $\dot{\chi}=0$, equation (\ref{result}) corresponds to the
Josephson current through a band-insulator with an energy gap $2|\Delta_c|$. Taking
the limit $|\Delta_c| \rightarrow 0$ reproduces the well-known expressions for
an ideal S/N/S junction \cite{zaitzev,beenakker}. In the range $\hbar \dot \chi
<|\Delta _s|<|\Delta _c|$ characteristic oscillations are expected in the
Josephson current as a function of the CDW velocity. 

Equation (\ref{result}) expresses the total current through the S/C/S junction as a 
function of the superconductor phases $\phi_{R,L}$ and the sliding velocity $\dot{\chi}$.
In order to obtain a closed set of equations, however, it is necessary to
derive an additional relationship between $\dot \chi $ and the
superconductor phase difference.
The additional closing relation can be derived microscopically from the
self-consistency relation for the phase of the CDW order parameter 
\begin{equation}
\int d\epsilon {\rm Tr}{\bbox{\Sigma}}_{3}\tilde{\bf g}^K=0,
\end{equation}
and the Keldysh functions in the reservoirs. The self-consistent solution for the
clean junction requires an electrochemical potential difference $\delta \mu
=\mu _R-\mu _L$ in the sliding state $\delta \mu =\hbar \dot \chi ,$ 
as in Ref. \onlinecite{rejaei}. This implies that below the critical
Josephson current the CDW will not move. Biased above the critical current
the CDW slides, but since the collective CDW motion is dissipative due to
the contact reservoirs, a small potential difference is induced on the
superconducting leads and the superconductor phase-difference will evolve
slowly in time. As a consequence we obtain the relation 
\begin{equation}
\label{result2}
\delta \mu =\hbar \dot \varphi /2=\hbar \dot \chi.
\end{equation}
The sliding CDW mode thus acts as a
shunting resistor parallel to the Josephson channel. The above
quasi-stationary approximation, where the dynamics of the superconductor phase is neglected
on the Josephson term, is therefore justified a posteriori \cite
{likharev}. In the short junction limit $L\ll \hbar v_F/|\Delta _s|$,
the dynamical phase $\dot \chi L/v_F$ can be neglected and if $|\Delta_c|\gg|\Delta_s|$ 
the Josephson current can be written as $I_c\sin\varphi$, where
$I_c$ is the critical current. The time-averaged potential becomes $<V>=R_Q\sqrt{I^2-I_c^2}$ and 
it oscillates with the Josephson frequency $\omega_{J} =2e<V>/\hbar $, where $R_Q=h/2e^2$ is the (spin included)
quantum resistance of a one-dimensional ballistic quantum wire. Note that far above the critical current $I\gg I_c$ the 
Josephson frequency becomes two times the fundamental frequency of the narrow band noise $\omega_J=2\omega_{NBN}=
2\pi I/e$.

The assumption of an ideal S/C/S junction is rather restrictive in realistic
systems where interface effects and impurities tend to pin the CDW. Although a fully
microscopic treatment is beyond present work, one can account for the pinning
effects by replacing Eq. (\ref{result2}) by $\delta\mu=\hbar{\dot{\varphi}}/2=
\hbar\dot{\chi}+\mu_{T}\sin\chi$. This model corresponds to the 
single-particle model of CDW's \cite{gruner}, where $\mu_{T}$ represents the threshold
value required to overcome the pinning potential.

We conclude by summarizing our results. We have formulated the kinetic
equations for a coupled superconductor-CDW system and calculated the
current through an ideal superconductor/charge density wave/superconductor junction. 
In the dc limit the CDW does not move. Biased above the critical current the CDW slides,
and the Josephson current
oscillates as a function of the CDW velocity. The collective mode acts as a
non-linear shunting resistance parallel to the Josephson channel. 

This work is part of the research program of the "Stichting voor
Fundamenteel Onderzoek der Materie (FOM)", which is financially supported by
the "Nederlandse Organisatie voor Wetenschappelijk Onderzoek (NWO)." We are
indebted to Yuli Nazarov and Gerrit Bauer for their valuable advise and
insights.

\begin{figure}
\caption{Schematic figure of a charge density wave with length $L$, sandwiched between
two superconductors with phases $\phi_{R}$ and $\phi_{L}$. The junction is biased
with a current source.}
\label{fig:fig1}
\end{figure}


\begin{references}
\bibitem{josephson}  B.D. Josephson, Physics Letters {\bf 1}, 251 (1962).

\bibitem{kulik}  I.O. Kulik, Zh. Eksp. Teor. Fiz. {\bf 57}, 1745 (1969)
[Sov. Phys. JETP {\bf 30}, 944 (1970)].

\bibitem{odintsov}  R.Fazio, F.W.J. Hekking, and A.A. Odintsov, Phys. Rev.
Lett. {\bf 74}, 1843 (1995).

\bibitem{gorkov}  For a review see {\it Charge Density Waves in Solids},
edited by L.P. Gor'kov and G. Gr\"uner (North-Holland, Amsterdam, 1989).

\bibitem{zant}  H.S.J. van der Zant, O.C. Mantel, C. Dekker, J.E. Mooij and
C. Traeholt, Appl. Phys. Lett. {\bf 68}, 3823 (1996).

\bibitem{larkin}  A.I. Larkin and O.V. Ovchinikov, Sov. Phys. JETP {\bf 41},
960 (1975) [Zh. Eksp. Teor. Fiz. {\bf 68}, 1915 (1975)].

\bibitem{artemenko}  S.A. Artemenko and V. Volkov, Sov. Phys. JETP {\bf 53},
1050 (1980) [Zh. Eksp. Teor. Fiz. {\bf 80}, 2018 (1981)].

\bibitem{zaitzev}  A.V. Zaitsev, Sov. Phys. JETP {\bf 59}, 1015 (1984) [Zh.
Eksp. Teor.Fiz. {\bf 86}, 1742 (1984)].

\bibitem{rejaei}  B. Rejaei and G.E.W. Bauer, Phys. Rev. B {\bf 54}, 8487
(1996).

\bibitem{zaitzev}  A.V. Zaitsev, in {\it Nonequilibrium Superconductivity},
edited by V.L. Ginzburg (Nova Science Publishers, New York, 1988).

\bibitem{likharev}  K.K. Likharev, in {\it Dynamics of Josephson Junctions
and Circuits} (Gordon an Breach Science Publishers, 1986).

\bibitem{beenakker}  C.W.J. Beenakker, in {\it Transport Phenomena in
Mesoscopic Systems}, edited by H. Fukuyama and T. Ando (Springer-Verlag,
Berlin, 1992).

\bibitem{visscher}  M.I. Visscher and B. Rejaei (unpublished).

\bibitem{gruner} G. Gr\"uner, A. Zawadowski, and P.M. Chaikin, Phys. Rev.
Lett. {\bf 46}, 511 (1981).

\end{references}
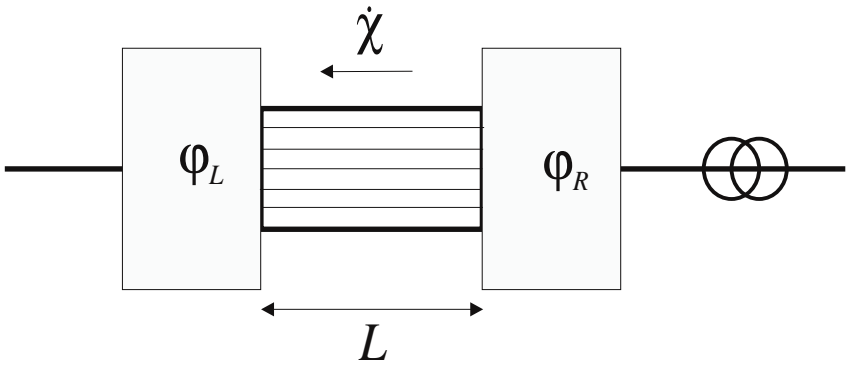
\end{document}